# Low-loss Broadband Negative Refractive Index due to Nonresonant 2D Helical Chiral Metamaterial

Kun Song,[1] Min Wang,[1] Zhaoxian Su,[1] Changlin Ding,[1] Yahong Liu,[1] Chunrong Luo,[1] Xiaopeng Zhao,[1] Khagendra Bhattarai,[2] & Jiangfeng Zhou[2]

[1]Department of Applied Physics, Northwestern Polytechnical University, Xi'an, 710129, China.
[2]Department of Physics, University of South Florida, 4202 East Fowler Ave, Tampa, FL, 33620-5700. Correspondence and requests for materials should be addressed to K.S. (email: songkun@nwpu.edu.cn) or J.Z. (email: jiangfengz@usf.edu)

In this paper, we demonstrated a 2D helical chiral metamaterial that exhibits broadband strong optical activity resulting nonresonant Drude-like response. The strong chirality leads to broadband negative refractive index with high figure of merit (>90) and extremely low loss (<2% per layer). The optical activity are insensitive to the angles of incident electromagnetic waves, thereby enabling more flexibility in polarization manipulation applicaitons.

**Introduction**

Manipulating the polarization states of electromagnetic waves has been a long-time interest in realms of life sciences, photoelectrons, telecommunications, *etc.*. Materials with chirality, which can rotate the polarization plane of electromagnetic waves, are very suitable for designing polarization converters. However, the chirality in natural materials is usually extremely weak and dispersive, the polarization converters made of natural materials are unrealistic because of huge thicknesses being much larger than the operating wavelengths or narrow operating bandwidths, especially at gigahertz and terahertz frequencies. Thus, materials that possess strong chirality are highly desired.

Metamaterials enable numerous of extraordinary electromagnetic phenomena that do not exist in natural materials, for instance, abnormal refraction or reflection[1-6], super-resolution imaging[7,8], cloaking[9-11], perfect absorption of electromagnetic waves[12-14]. The emerging of metamaterials makes it possible for us to obtain strong chirality. In the past few years, chiral metamaterials (CMMs) have been constructed to realize negative refractive index[15-20], strong optical activity[21-25], circular dichroism[26-30], as well as asymmetric transmission[31-35]. Using strong resonances, the optical rotatory power of CMMs might rise up to several orders of magnitudes larger than that of natural materials[36-40]. However, owing to the inherent Lorentz-like resonances, the giant optical activity of the previous CMMs is generally accompanied by high losses, high dispersion, narrow transmission bandwidth, and strong polarization distortion because of large ellipticity [18-20,36,41-43], which is exceedingly detrimental for designing broadband and efficient polarization rotators. Recently, there has been much efforts devoted to exploring nondispersive optical activity. By combining a meta-atom with its complement in a chiral configuration, low dispersive optical activity at the transmission resonance has been demonstrated; while this kind of CMMs are subjected to narrow transmission bandwidths[44-46]. In more recent papers, three-dimensional off-resonant or nonresonant types of CMMs were also demonstrated to achieve nondispersive optical activity[47,48]. However, the complicated fabrication process needed for creating these three-dimensional structures are very challenging even with current state-of-art technologies, especially at the optical part of the spectrum. Therefore, CMMs with simplified architectures that can be easily made using standard micro- or nano-fabrication processes are highly demanded.

In this paper, we propose a CMM working at GHz frequencies that is composed of a bi-layer helix structures with simple geometry where the interconnection between the bi-layer can be easily realized. Compared to previous work [48], our design be scale down to make CMMs operating in THz and even infrared regime utilizing standard micro- or nanofabrication techniques. Unlike the Lorentz-like resonances in the previous CMMs[18-20,36-38,41,42], the present CMM exhibits nonresonant Drude-like response due to the connection between two layers. Simulation, calculated, and experimental results show that our CMM exhibits strong nondispersive optical activity, high transmittance and extremely low ellipticity in a broadband frequency range. The optical activity of multilayer CMMs is proportional to the number of CMM layers. We have also found that the optical activity are independent of the incident angle and the dielectric constant of the substrate, which has not be reported previously to the best of our knowledge. Moreover, it is also found that the transmission frequency of the proposed CMM presents dynamical tunability by altering the permittivity of dielectric substrate, indicating that the CMM is very suitable for designing frequency-tunable polarization manipulation devices for telecommunication applications.

## Results

**Design of unit cells and theoretical calculation.** Figure 1(a) and,1(b) show the schematic view and photograph of the present CMM, respectively. As shown in Fig. 1(a), the unit cell is composed of four interconnected double-helix structure. Along the clockwise direction, each double-helix are rotated by $\pi/2$ from its adjacent neighbor. Array of unit cells forms a C4 square lattice with four-fold symmetry which ensures a pure optical activity effect. The metal slices on the front and back surfaces of the dielectric substrate are connected via metallization holes. Obviously, all the unit cells of the designed CMM are interconnected. The geometrical parameters of the unit cells are as follows: $a = 3$ mm, $b = 1.3$ mm, $d = 0.6$ mm, $g = 0.3$ mm, $l = 1.7$ mm, $r = 1.5$ mm, $t = 3.07$ mm, and $w = 0.9$ mm. The metal copper cladding is 0.035 mm in thickness with a conductivity of $\sigma = 5.8\times10^7$ S/m. The permittivity of the F4BM-2 substrate is $2.65+0.001*i$. In the experiments, a CMM sample composed of 40*40 unit cells was fabricated via the printed circuit board etching process.

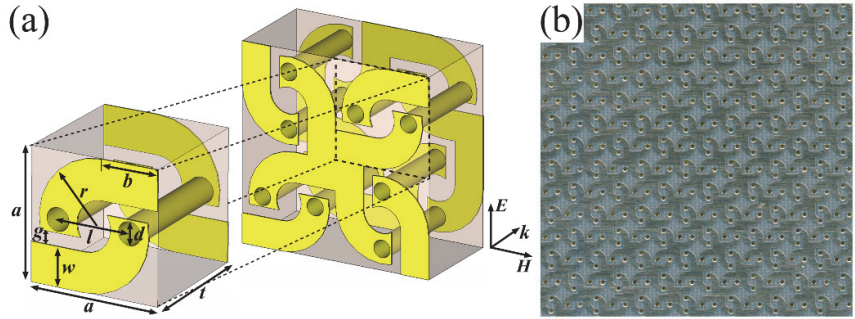

**Figure 1. Design of CMM** (**a**) Schematic diagram and (**b**) photograph of the proposed CMM.

To get the electromagnetic properties of the present CMM theoretically, we use the effective current model to calculate the effective parameters of the CMM[48]. Due to the interconnected metal structures, the current can flow freely in the CMM, which will lead to a Drude-like response. For the double helix structure, the inductance and resistance are approximatively calculated as $L = \mu_0 R(\ln\frac{8R}{w} - 2)$ and $R_\Omega = \sqrt{\frac{\mu_0 \omega S}{2\sigma w^2}}$, respectively, where $R = \sqrt{S} = \sqrt{(l+d)t}$, $S$ is the effective cross section area of the double helix structure[48-50] The electric potential of time-harmonic electromagnetic field can be expressed as:

$$\xi = aE - (\pm i\omega\mu_0 SH) = (R_\Omega - i\omega L)I . \tag{1}$$

Here, the signs $\pm$ represent the right-handed and left-handed helix structure; $I$ is the total current flowing in the helix structure. From equation (1), we can obtain:

$$I = \frac{aE}{R_\Omega - i\omega L} - \frac{\pm i\omega\mu_0 SH}{R_\Omega - i\omega L} . \tag{2}$$

Then, the electric and magnetic dipole moments can be obtained via the following formulas[48]:

$$\boldsymbol{P} = \frac{\boldsymbol{J}}{-i\omega} = \frac{a\boldsymbol{I}}{-i\omega V} = -\frac{a^2}{(i\alpha\omega + \omega^2)LV}\boldsymbol{E} - \frac{\pm\mu_0 aS}{(\alpha - i\omega)LV}\boldsymbol{H} , \tag{3}$$

$$\boldsymbol{M} = \pm\frac{\boldsymbol{IS}}{V} = \frac{\pm aS}{(\alpha - i\omega)LV}\boldsymbol{E} - \frac{i\omega\mu_0 aS^2}{(\alpha - i\omega)LV}\boldsymbol{H} , \tag{4}$$

Where $V$ is the volume of the unit cell and $\alpha = \frac{R_\Omega}{L}$ is the dissipation constant. From equations (3) and (4), the electric displacement vector $\boldsymbol{D}$ and magnetic flux density vector $\boldsymbol{B}$ can be expressed as follows:

$$\boldsymbol{D} = \varepsilon_0 \boldsymbol{E} + \boldsymbol{P} = [\varepsilon_0 - \frac{a^2}{(i\alpha\omega + \omega^2)LV}]\boldsymbol{E} - \frac{\pm\mu_0 aS}{(\alpha - i\omega)LV}\boldsymbol{H} , \tag{5}$$

$$\boldsymbol{B} = \mu_0(\boldsymbol{H} + \boldsymbol{M}) = \frac{\pm\mu_0 aS}{(\alpha - i\omega)LV}\boldsymbol{E} + \mu_0 \boldsymbol{H} - \frac{i\omega\mu_0^2 aS^2}{(\alpha - i\omega)LV}\boldsymbol{H} . \tag{6}$$

For the chiral media, the constitutive equation is as follows:

$$\begin{pmatrix} \boldsymbol{D} \\ \boldsymbol{B} \end{pmatrix} = \begin{pmatrix} \varepsilon_0 \varepsilon & -i\kappa c \\ i\kappa c & \mu_0 \mu \end{pmatrix} \begin{pmatrix} \boldsymbol{E} \\ \boldsymbol{H} \end{pmatrix}, \tag{7}$$

Where $\kappa$ is the chirality parameter. Using equations (5)-(7), the effective parameters of the proposed CMM can be derived as follows:

$$\varepsilon = \varepsilon_f - \frac{a^2}{(\omega^2 + i\omega\alpha)\varepsilon_0 LV} \tag{8}$$

$$\mu=\mu_f - \frac{\omega\mu_0 S^2}{(\omega+i\alpha)LV} \tag{9}$$

$$\kappa=\pm\frac{\mu_0 caS}{(\omega+i\alpha)LV} \tag{10}$$

Where $\varepsilon_f$ and $\mu_f$ are the quantitative fitting parameters. The transmission spectrum of the CMM is expressed in the terms of the following equation[48]:

$$T=\frac{4Ze^{ink_0h}}{(1+Z)^2-(1+Z)^2 e^{2ink_0h}}. \tag{11}$$

Here, $Z=\sqrt{\mu/\varepsilon}$ is the effective impedance; $n=\sqrt{\mu\varepsilon}$ is the effective average refractive index; $k_0=\omega\sqrt{\mu_0\varepsilon_0}$ is the wave number in vacuum. And the polarization azimuth rotation angle can be calculated as:

$$\theta=\frac{\mathrm{Re}(\kappa)\omega h}{c}. \tag{12}$$

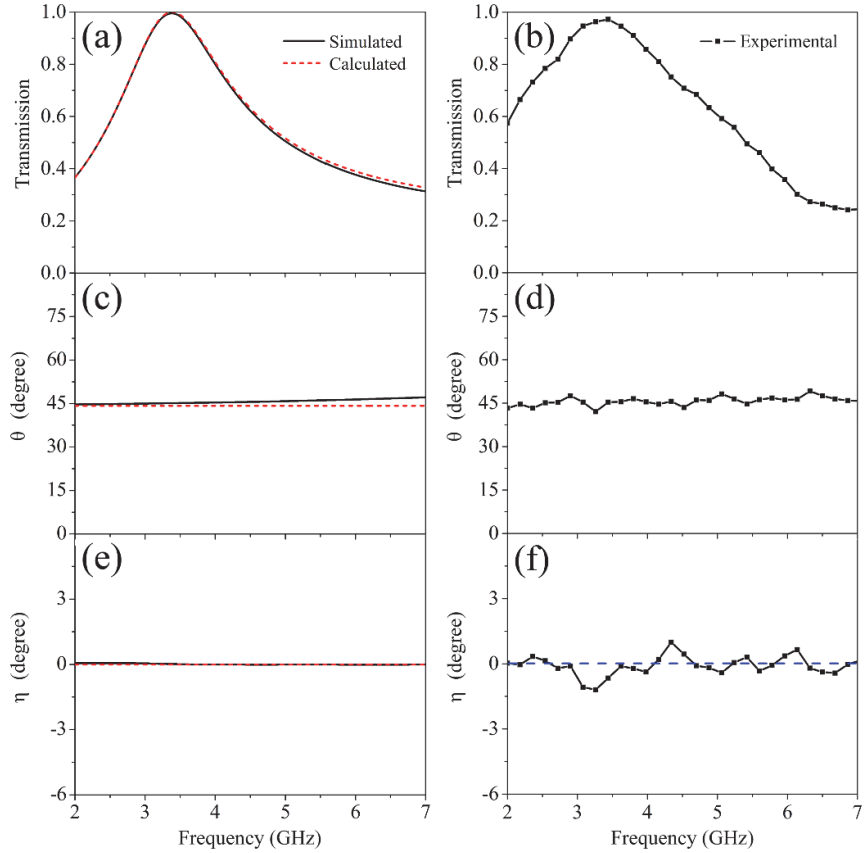

**Figure 2 Transmission and optical Activity.** The simulated, calculated, and experimental results

of the single-layer CMM at normal incidence. (**a**) and (**b**) Transmission spectra ($T = \sqrt{T_{xx}^2 + T_{yx}^2}$), (**c**) and (**d**) Polarization azimuth rotation angle $\theta$, (**e**) and (**f**) Ellipticity $\eta$. In the calculation, the quantitative fitting parameters $\varepsilon_f$ and $\mu_f$ are chosen as 21.3 and 0.95, respectively.

**Results of single-layer CMM.** Figure 2 shows the simulated, calculated and measured results of the single-layer CMM at normal incidence. We see clearly that the experimental results are in good qualitative agreement with the simulated and calculated ones Fig. 2(a, b) show that transmission reach to the peak value of nearly unity at 3.4 GHz. Most importantly, the transmittance is over 0.8 in the wide frequency range of 2.9 ~ 4.0 GHz with a relative bandwidth of 32%, indicating that the present CMM can achieve broadband and high-efficiency transmission. The polarization azimuth rotation angle of the CMM are plotted in Fig. 2(c,d). It is obvious from the figure that the polarization azimuth rotation angle is kept almost constant at about 45° within the entire measured frequency region, indicating that the CMM can realize broadband and nondispersive optical activity. Figure 2(e,f) depict the ellipticity $\eta$ produced by our structure . As we know, the chiral media will generate a pure optical activity effect at $\eta = 0$, *i.e.*, when a linearly polarized incident wave passes through the chiral media, the transmitted wave remains linearly polarized but with the polarization plane rotated by an angle of $\theta$[19]. Since the values of ellipticity in Fig. 2(e,f) are nearly zero in the whole frequency range, the transmission spectrum is still linearly polarized without distortion of polarization states.

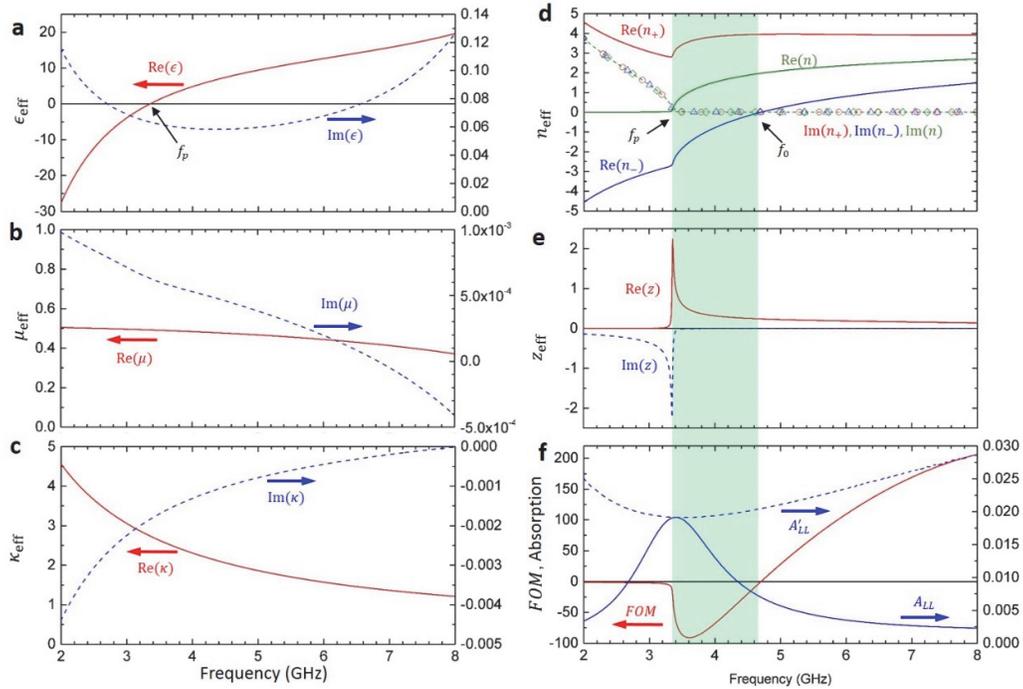

**Figure 3 Effective medium parameters.** Effective permittivity (**a**), effective permeability (**b**), effective chirality (**c**), effective refractive indices (**d**), effective impedance (**e**) of CMM calculated

from simulated transmission and reflection coefficients of circular polarizations. Red circle, blue triangle and green diamond with dashed curves in (**d**) show the imaginary parts of $n_+$, $n_-$ and $n$, respectively. (**f**) Figure of merit of $n_{LL}$ and absorption for left-handed circular polarized light.

Our CMM exhibits four-fold symmetry and shows pure chirality with zero cross-polarization transmissions ($T_{RL} = T_{LR} = 0$), thereby can be described as a bi-isotropic medium. To reveal the mechanism of the giant nonresonant optical activity, we calculated the effective parameters using a retrieval method based on effective bi-isotropic medium. In our calculation, we added 0.5 mm air space at both side of the CMM, so the thickness of the effective medium is $d = t + 1$ mm = 4.07mm. As shown in Fig. 3a, the effective permittivity exhibits a Drude-like response, where the real part, $\text{Re}(\epsilon_{\text{eff}})$, increases from negative values to positive values as frequency increases. This arises from the continuity of metallic mesh in the 2D plane perpendicular to the wave propagation direction. The plasma frequency, $f_p = 3.35$ GHz (for $\text{Re}(\epsilon_{\text{eff}}) = 0$), can be tuned by the density of the metallic mesh through geometric parameters $a$ and $t$. The effective permeability in Fig. 3b remains nearly constant, $\text{Re}(\mu_{\text{eff}}) \approx 0.5$ showing a diamagnetic behavior. The large chirality, $\text{Re}(\kappa_{\text{eff}})$, shows a nonresonant and slow varying dispersion as shown in Fig. 3c, leading to nearly constant giant azimuth rotation $\theta = \text{Re}(\kappa_{\text{eff}}) \cdot k_0 d \approx 45°$. The imaginary part, $|\text{Im}(\kappa_{\text{eff}})|$, is less than $1 \times 10^{-3}$ over the entire frequency range from 2 GHz to 8 GHz, resulting in nearly zero ellipticity $\eta = \text{Im}(\kappa_{\text{eff}}) \cdot k_0 d$. Note that the imaginary parts of $\epsilon$, $\mu$ and $\kappa$ are extremely small compared to their real parts, indicating very low loss. Figure 3d shows the effective refractive indices, $n_\pm = n \pm \kappa$, for RCP ($n_+$) and LCP ($n_-$) incident waves, respectively. Note that for bi-isotropic medium, the eigen solutions of wave equations are LCP and RCP waves with refractive indices defined as $n_\pm = k_\pm/k_0 = n \pm \kappa$, where $k_+$, $k_-$, are wave numbers for RCP and LCP waves, respectively, and $k_0 = 2\pi/\lambda$ is the wave number in vacuum. The regularly defined refractive index in isotropic medium, $n = \sqrt{\epsilon\mu}$, does not cause any physical effect on the propagation of waves in bi-isotropic medium. Here we only use $n$ helping us to understand the effects of $\epsilon$ and $\mu$. In Fig. 3d, negative values of $\text{Re}(\epsilon_{\text{eff}})$ below plasma frequency ($f_p$=3.35 GHz) leads to purely imaginary $n_{\text{eff}}$ (green curve) representing a typical plasma medium. Above $f_p$, $n_{\text{eff}}$ shows a low-loss dielectric medium with positive $\text{Re}(n)$ and nearly zero $\text{Im}(n)$. Because of the chirality, the effective refractive index of CMM split into two branches for LCP ($n_-$) and RCP ($n_+$) incident waves, respectively. Importantly, for frequency below $f_0 = 4.70$ GHz, we observed $\text{Re}(n) < \text{Re}(\kappa)$, leading to negative refractive index for LCP wave (i.e. $\text{Re}(n_-) < 0$). Unlike negative index within a narrow band around the resonance frequencies of the twisted double-layer resonators, our CMM shows ultra-wide bandwidth from 4.7 GHz down to DC. In particular, in the range between $f_p$ and $f_0$ (green highlighted region in Fig. 3d), $n_-$ has extremely small imaginary part, and thereby exhibiting very high figure of merit, $FOM = \text{Re}(n)/\text{Im}(n)$. As shown in Fig. 1f, FOM in this region reaches maximum value of 91, which is significantly (30 times) higher than other low-loss negative index metamaterials with FOM<3. Such high FOM arises from extremely low loss of nonresonant Drude-like responses. As shown in Fig. 3f, the absorption of LCP wave, $A_{LL} = 1 - |T_{LL}|^2 - |R_{LL}|^2$ reaches a peak of 0.019 at $f_p$=3.38 GHz where the transmission (Fig. 2a) also closes to its maximum. This level of absorption is extremely small compared to previous negative index metamaterials (A~0.5) due to resonant $\mu_{\text{eff}}$ and/or $\epsilon_{\text{eff}}$. Note that the peak of $A_{LL}$ in the green highlighted region is due to better impedance matching than other frequencies. As shown in Fig. 3e, the effective impedance, $z_{\text{eff}}$, equals to 1 at frequency 3.40 GHz, matching with the impedance of sounding air space ($z_0 = 1$). This leads to the nearly 100% transmission as shown in Fig. 2a. As frequency moves away, $z_{\text{eff}}$ rapidly decreases to nearly zero, thereby causing increase of reflection. The normalized absorption, $A'_{LL} = A_{LL}/(1 - |R_{LL}|^2)$, calculates the absorption of wave energy that enters the metamaterial, thereby better describing the loss. As shown in Fig. 3f, $A'_{LL}$ reaches minimum values in the green highlight region where FOM is high.

The electromagnetic properties of the single-layer CMM at oblique incidence are shown in Figure 4. The incident angle of electromagnetic wave is increased by the step of 10°. Figure 4(a,b) show the simulation and experimental transmission spectra under different incident angles. It shows that, although the transmission peak generates a very slight blue shift simultaneously accompanied by a slight reduction of the maximum transmittance as the incident angle increases, the broadband and efficient transmission performance still exists. In Fig. 4(c,d), it is of significance that the polarization azimuth rotation angle $\theta$ in the whole frequency region is kept approximately constant at 45° with the incident angle increasing from 0° to 40°, which implies that the single-layer CMM can realize nondispersive optical activity in a wide region of incident angle. As shown in Fig. 4(e,f), the ellipticity of the single-layer CMM gradually increases with response to the increment of the incident angle. However, the maximum value of the ellipticity is less than 1.0° even if the incident angle rises up to 40°. As the ellipticity is very small, the transmission spectrum of the single-layer CMM can still be regarded as linearly polarized. These facts reveal that the broadband high transmission and nondispersive optical activity behaviors of the single-layer CMM can be maintained regardless of the incident angles, exhibiting more advantages than the CMMs previously reported[18-20,36-38,41,42].

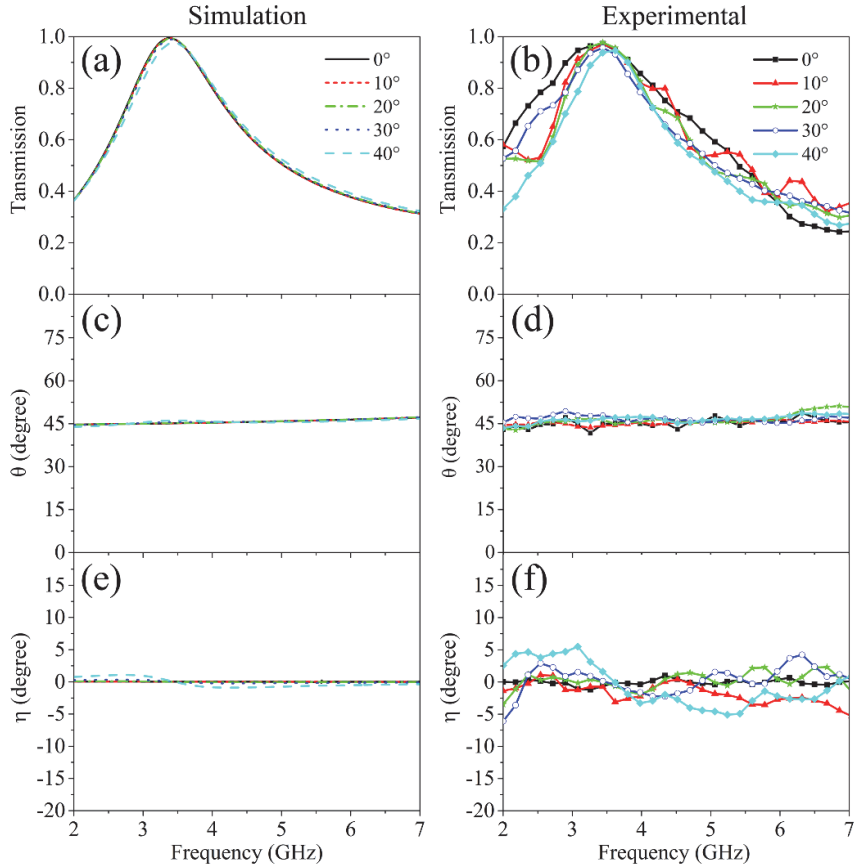

**Figure 4 Oblique incidence.** The simulation (left column) and experimental (right column) results of the single-layer CMM at oblique incidence. (**a**) and (**b**) Transmission spectra, (**c**) and (**d**) Polarization azimuth rotation angle $\theta$, (**e**) and (**f**) Ellipticity $\eta$.

**Results of dual-layer CMMs.** The numerical and measured results of the dual-layer CMMs with different interlayer spacings at normal incidence are shown in Fig. 5. It is seen in Fig. 5(a,b) that there are two transmission peaks occurring on the transmission spectra for the dual-layer CMMs. As the interlayer spacing increases, the two transmission peaks exhibit significant red shifts and come close to each other. These phenomena arise due to the double layers of CMMs, that form a Fabry-Perot-like resonant cavity. The Fabry-Perot-like resonance will occur when the electromagnetic waves are reflected between the double layers of CMMs, resulting in the generation of two transmission peaks. Moreover, the Fabry-Perot-like resonant cavity with different space distances will generate different resonant responses, which leads to the transmission spectra altering with the interlayer spacings[48].

Figure 5(c-f) portray the polarization azimuth rotation angle and ellipticity of dual-layer CMMs evolving with different interlayer spacings, respectively. It is obvious that the polarization azimuth rotation angle and ellipticity within the whole frequency region has been kept approximatively at 90° and 0°, respectively, and the alteration of the interlayer spacing has almost no influence on the optical activity and ellipticity of the dual-layer CMMs. The fascinating properties mentioned above imply that the dual-layer CMMs can function as a 90° polarization rotator, of which the transmission spectrum can be dynamically tuned by varying the space distance between the two layers of CMMs. Furthermore, compared Fig. 5(c,d) with Fig. 2(c,d), it can be found that the optical activity of dual-layer CMMs is just two times of that of the single-layer CMM. More simulations demonstrate that the optical activity of multi-layer CMMs is proportionl to the number of CMM layers (see Fig. S5 in Supplementary).

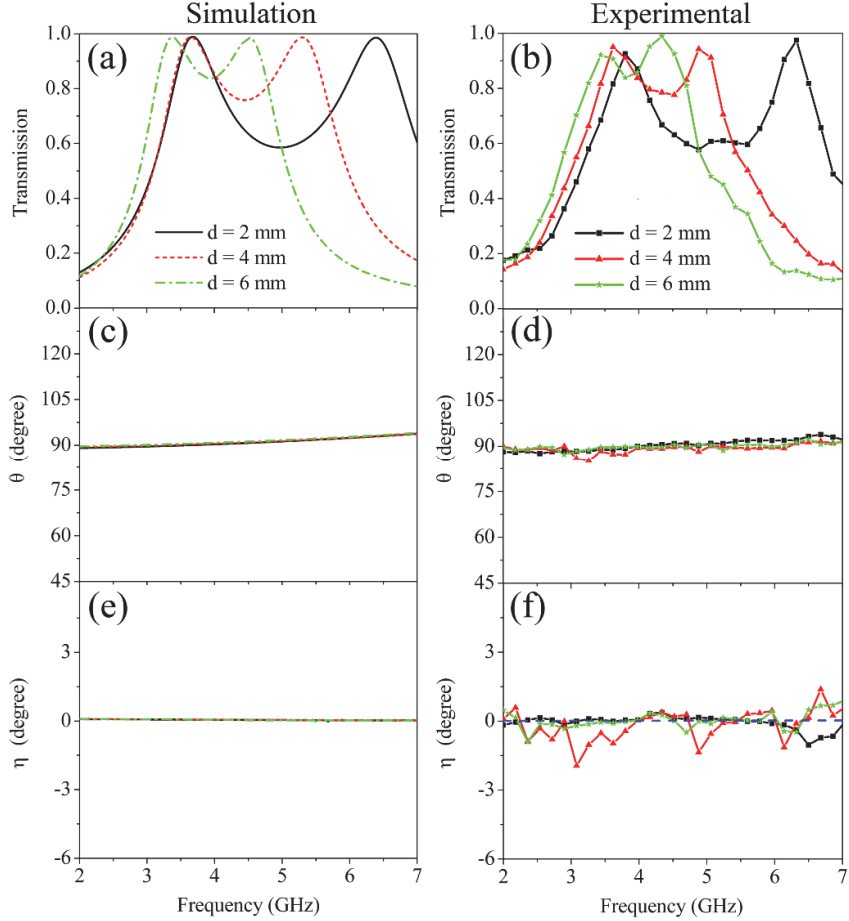

**Figure 5 Dual-Layer CMM.** The numerical (left column) and measured (right column) results of the dual-layer CMM with different interlayer spacings at normal incidence. (**a**) and (**b**) Transmission spectra, (**c**) and (**d**) Polarization azimuth rotation angle $\theta$, (**e**) and (**f**) Ellipticity $\eta$.

## Discussion

In order to get physical insight into the mechanism of broadband nondispersive optical activity and zero ellipticity, we examine the surface current distributions of the considered CMM, as shown in Fig. 6(a). It is significant that the current flows freely in the interconnected metal patterns. Consequently, the CMM exhibits a nonresonant Drude-like response, which can effectively avoid the dispersive optical activity due to the Lorenz-like resonances existing in the previous CMMs[16,19-21]. Figure 6(b) presents the schematic illustration of the flowing direction of surface current. It is seen that the free-flowing current can induce the magnetic moments ***M*** being in the opposite direction to the external electric field ***E***. And in Fig. 6(c) the simulated magnetic field distribution in the middle plane of the unit cell also confirms the generation of the induced magnetic moments ***M***. These results imply that strong cross-coupling effect occurs as the electromagnetic waves pass through the CMM. This is the origin of the chirality of the proposed CMM. Figure 6(d) plots the numerical results of Re($\kappa$), Im($\kappa$), and Re($\kappa$)*$\omega$/2$\pi$. Our calculated result confirms that the dissipation constant $\alpha$ in equation (10) is far smaller than the angular frequency $\omega$. As the chirality parameter $\kappa$ is a finite value, the imaginary part of $\kappa$ is thus nearly zero and can be ignored. Then, the equation (10) can be approximatively expressed as $\kappa \approx \pm \frac{\mu_0 caS}{\omega LV} \approx \text{Re}(\kappa)$, from which we can found that the real part of $\kappa$ is inversely proportional to $\omega$. As a result, the value of $\text{Re}(\kappa)*\omega$ is a constant. Finally, the polarization azimuth rotation angle $\theta$ will be constant in the whole frequency range according to equation (12), *i.e.*, the nondispersive optical activity occurs. Additionally, the circular dichroism of CMMs is closely related to the imaginary part of $\kappa^{48}$. Since imaginary part of $\kappa$ is approximatively zero, the circular dichroism is therefore absent, which results in the ellipticity being zero.

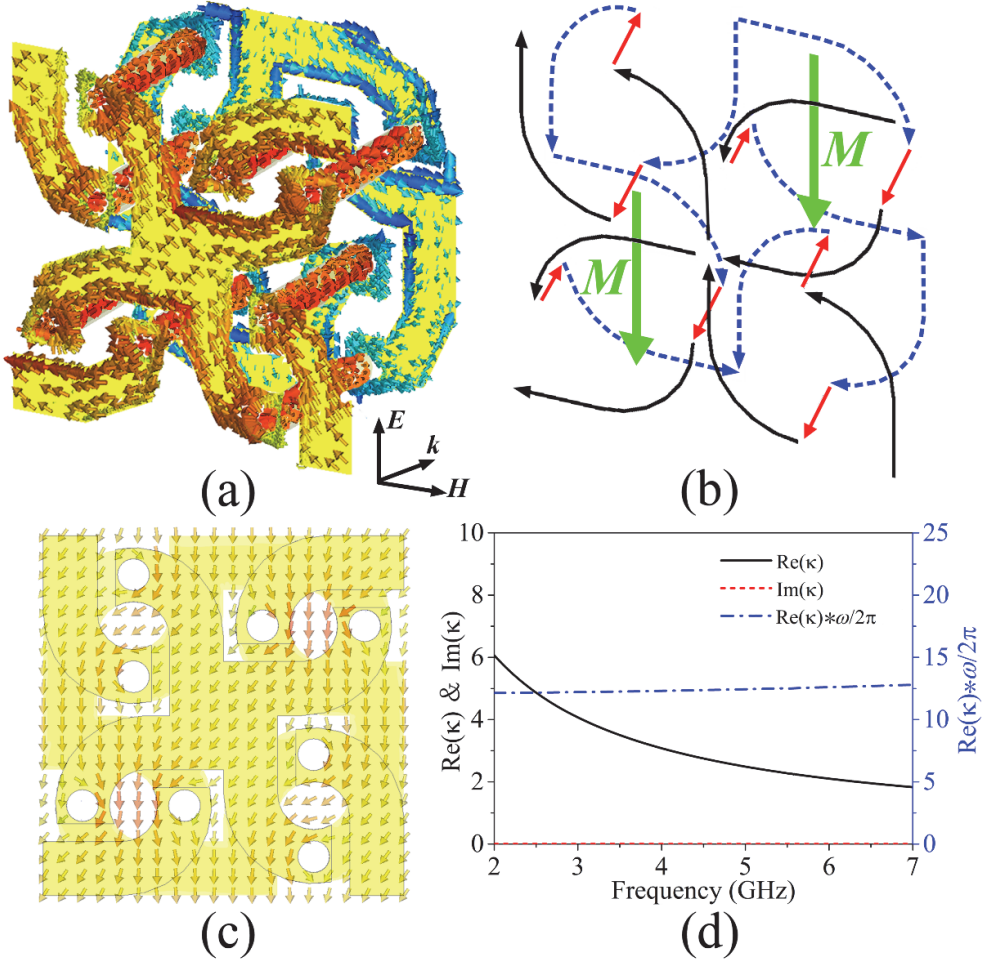

**Figure 6 Current Distribution.** (**a**) Surface current distribution of the designed CMM. (**b**) Schematic illustration of the flowing direction of surface current. (**c**) The simulated ***H*** field distribution at the middle plane of the unit cell. (**d**) The numerically calculated results of Re($\kappa$), Im($\kappa$), and Re($\kappa$)*$\omega$/2$\pi$.

**Conclusions**

In summary, we have demonstrated a fascinating CMM that consists of double helical structures. As a result of the interconnected metal structures, this kind of CMM generates Drude-like response, which combines with the C4 symmetry geometry giving rise to the broadband nondispersive optical activity and zero ellipticity simultaneously accompanied by high transmittivity. Most notably, the broadband behaviors of nondispersive optical activity and high transmission of the single-layer CMM are independent of the incident angles of electromagnetic waves and permittivity of dielectric substrate. And the transmission spectrum of the CMM can be successively tuned by varying the permittivity of substrate but with the optical activity unchanged. In addition, the optical activity of multi-layer CMMs is proportional to the number of CMM layers without being affected by the interlayer coupling effect, which enables us to obtain much stronger optical activity just by simply increasing the layer number. With the intriguing properties, the elaborate CMM exhibits more application flexibility and is greatly appealing for controlling the polarization state of electromagnetic waves.

## Methods

**Numerical simulation.** Simulations were performed with the commercial software CST Microwave Studio. In the simulations, a linearly polarized wave was incident on the sample; the unit cell boundary conditions were employed in the $x$ and $y$ directions and open boundary conditions were utilized in the $z$ direction.

**Experimental measurement**. The experimental measurements were carried out by using an AV3629 network analyzer with two broadband linearly polarized horn antennas in an anechoic chamber. Using the linear co-polarization and cross-polarization transmission coefficients $T_{xx}$ and $T_{yx}$, we can obtain the transmission coefficients of the circularly polarized waves by the formula $T_{\pm} = T_{xx} \mp i * T_{yx}$, where $T_{+}$ and $T_{-}$ are the transmission coefficients of the right-handed and left-handed circularly polarized waves, respectively. The optical activity is revealed via the polarization azimuth rotation angle $\theta = \dfrac{\arg(T_{+}) - \arg(T_{-})}{2}$. And the circular dichroism of the transmitted waves is characterized by the ellipticity $\eta = \dfrac{1}{2}\arctan\dfrac{|T_{+}|^2 - |T_{-}|^2}{|T_{+}|^2 + |T_{-}|^2}$.

## Acknowledgements

This work was supported by the National Natural Science Foundation of China (Grant Nos. 11174234, 11204241, 11404261, 61601375, 61601367), the Fundamental Research Funds for the Central Universities (Grant No. 3102016ZY029, 3102016ZY028) and the Northwestern Polytechnical University Scientific Research Allowance (Grant No. G2015KY0302). The USF portion of this work was supported by the Alfred P. Sloan Research Fellow grant BR2013-123 and by KRISS grant GP2016-034.

## Author Contributions

K.S. and J.Z. conceived the idea, designed the experiment, and supervised the project. K.S., M.W. and Z.S. performed the numerical simulations. Y.L. and C.D. fabricated the sample and carried out the experimental measurements. K.S., C.L., X.Z., K.B. and J.Z. did the theoretical analysis. K.S. and J.Z. co-wrote the manuscript. All the authors have reviewed the manuscript.